\documentclass[sigconf,11pt]{acmart}
\usepackage[export]{adjustbox}
\renewcommand\footnotetextcopyrightpermission[1]{}

\AtBeginDocument{%
  \providecommand\BibTeX{{%
    \normalfont B\kern-0.5em{\scshape i\kern-0.25em b}\kern-0.8em\TeX}}}

\setcopyright{none}
\copyrightyear{}
\acmYear{}
\acmDOI{}

\acmConference[CSE6242]{Make sure to enter the correct
  conference title from your rights confirmation emai}{Sept 01,
  2022}{Atlanta, GA}
\acmPrice{}
\acmISBN{}

\begin{document}

\title{\Large EXPLORE - Explainable Song Recommendation}

\author{\normalsize Abhinav Arun, Mehul Soni,  Palash Choudhary, Saksham Arora}
\renewcommand{\shortauthors}{Trovato and Tobin, et al.}



\maketitle
\pagestyle{plain}


\section{Introduction}
We are conducting our project on building a music recommendation product with key focus on new song discovery and explainability. The system is an attempt to generate a playlist by suggesting songs heard by users having similar preferences. The product will take Spotify user-id of the user as an input and will generate a music recommendation playlist. The UI will also have toggles to select a mood to update the recommended playlist and will also have an explanation window for why a song was recommended. We plan to use a hybrid approach- a combination of collaborative filtering and content-based filtering- to develop our recommendation system. The user data will be procured from Spotify for Developers’ Web API and the user-song data is taken from subset of MLHD\cite{mlhd} dataset. The Spotify API provides access to multi-level data such as user’s playlist data, song’s attribute data, artist data etc and the MLHD data has user and song interaction data.

\section{Problem Definition}
There are state of the art song recommendation systems available such as Spotify, Apple Music, Amazon Music etc. but they lack in 2 key areas: 1) They fail to provide any explainability on why a  particular song was recommended. 2) These recommendation system do not provide users with tools to control the recommendations .
Through our product, we aim to address these challenges.

\section{Literature Review}
Several research papers have been published over the years on recommendation systems, and visualization in the context of song recommendation. A few of them are discussed below.

\subsection{Single User Recommendation}
Recommendation systems are widely used applications of machine learning. They provide suggestions for items that may be useful to the customer and help users avoid selection paralysis. As suggested in \cite{Masthoff2011} and \cite{ADIYANSJAH201999}, recommendation systems majorly use two approaches: i) Collaborative filtering i.e. assumption that people with similar preferences in past are likely to prefer same items in the future, and ii) content-based filtering i.e. recommendation based on the item’s features.
\vspace{5pt}

Although novel techniques for recommendation have been used in the domain of Deep Learning \cite{lee2022deep}, \cite{AdvancedRecommendation}, they lack in explainability and in performance with respect to sparse matrices.
Recommendations can be made explainable by using a variant of matrix factorization algorithm (AMCF -> Attentive Multitask Collaborative filtering)\cite{ConstrainedMatrix} that accounts for explainable recommendations \cite {DBLP:journals/corr/abs-1711-10816}. The model will learn to map the difficult-to-interpret latent features onto the space of interpretable input features space using an attention mechanism as used in \cite{explaininterpret}. Sparse autoencoders \cite{AHMADIAN2022116697} can also be used with a similar setup to provide better personalized recommendations.
We plan to perform user studies for evaluation as conducted in \cite{ExplainableSurveyEvaluation}.

\subsection{Group Recommendation*}
A key aspect is to identify how to make a recommendation for the group. \cite{Masthoff2011} suggests two methodologies: i) developing recommendations for individuals and aggregating it for a group, and ii) aggregating individual preferences into a group and using this group to make recommendations. Combining individual ratings of users we can aggregate them using different metrics such as Plurality Voting, Average, Least Misery, Most Pleasure and feed into the single user recommendation model.
 
 Designing explanations for group recommendations is another challenge\cite{grouprecommendation} This paper, in the case of Collaborative Filtering talks about providing explanations based on based on the ratings of the nearest neighbors. Simplistic strategies have also been suggested for content-based recommendation systems. 
 
 Few considerations while designing group recommendation systems is handling fairness, this paper\cite{grouprecommendation} talks about how including at least x items for each user can be one such way. Building Consensus is another nuance. It can be measured by checking the amount of disagreement between ratings of users.
 
\subsection{Visualization}
We surveyed papers to identify effective method of user engagement through interaction. \cite{surveymusic} inspected the relationship between musicology and visualization. It's extensive task based analysis on previous existing visualizations provided insight towards our proposed visual interface. \cite{Simplealgorithms} discussed multiple algorithms to visualize network graphs such as forced directed network graph, network matrix etc.

Second, we look at the \cite{PersonalRec}, which attempted to tackle the problem of music exploration. A subsequent work \cite{MoodExplore} implemented a comparative study on the visualization experience of the previous paper and their own; the visualizations with user control helped justify recommendations. Therefore, we plan to incorporate mood control similar to \cite{MoodExplore}  Additionally, giving the users control over recommendations can be incorporated by providing mood based toggles\cite{Knowingmeknowingyou}. 

\section{Methodology}
The design of the product is discussed in below sections:
\subsection{Data Preparation}
We are using MLHD data set for building our recommendation system. The dataset has user-song interaction information of 992 users for 1083471 songs. Although the data set has records from 2006-2013, it is sparsely populated from 2009-2013 and thus we will only use 2006-2008 data for training our model. This data contains time based user-song interaction information. We will utilize this information to create a user-song interaction rating matrix. Rating is a score in range 1-5 that suggests how much a user likes the song. We developed this rating by calculating monthly song frequency and inverse song frequency. Monthly song frequency(TF) is defined as the number of times a user has listened to a song in a month and inverse song frequency(IDF) is defined as a function of the number of users who have listened to that song in that particular month. So overall, the rating for that month is given as:
$$\textbf{Rating} = \mathbf{TF_{Mt}}*\mathbf{IDF_{Mt}} $$\\
where,
$$\mathbf{IDF_{Mt} = \log(\frac{N}{1+df})}$$\\
N : Total Number of users\\
df : Number of users who have listened to that song in that particular month\\
This is done so that we capture niche user tastes in our song recommendation and songs that are listened by relatively less users is given higher weightage. Further, to capture temporal relationship, we give relatively higher weightage to recent songs i.e. most recent month is given weight as 1, second recent month is given weight as (23/24) and so on.

Further, we extracted following data-sets from Spotify's Web API: 1) New user's song data, 2) MLHD Song's attributes, and 3) Top 500 songs of 2022 and Top 500 songs of All time playlist's songs and their attributes.

\subsection{Recommendation Algorithms}
We will be using the following set of algorithms for forking out recommendations to the end users : 
\vspace{5pt}
\\\\\
$\textbf{Collaborative Filtering}$\\ In this technique, we can base our algorithm either on user-user or on song-song similarity. We identify a set of closest neighbors for a given user i (identified through the Pearson correlation coefficient) based on their ratings for common songs. Then, we take the weighted average of the ratings that these neighboring users give to a song j in order to come up with a predicted rating r(i,j) for a given user i for the song j. Also , to account for user bias, we compute the same on deviations of ratings around the mean for a given user.\\
 \\
 $\textbf{Matrix Factorization Algorithm}$\\
 This algorithm uses the concept of singular value decomposition technique to express users and songs in terms of latent factors (k-dimensional latent feature space). We can then use cosine distance  to identify the songs with the highest similarity to a given user's liking.\\
 \\
 $\textbf{Explainable Matrix Factorization Algorithm}$\\
 This algorithm is a variant of the above-mentioned matrix factorization algorithm wherein we also provide as input , external metadata in the form of interpretable song features which could be used to help explain recommendations. For each item latent factor j, we train a predictor $$f_{j} \approx i_{ij}$$ in terms of the known interpretable attributes for a song i. Thus, we would get a mapping from the interpretable feature to latent feature set .

 This mapping would help us understand which of the interpretable features are driving the individual latent factors.

Now, given that a song (s) was recommended to a user (u) , it means that the feature vectors of the song and user in latent space were similar. Using this, we can identify the latent dimension in which the feature vectors had the highest similarity. Given this latent dimension, we can use the above-computed mapper to get the corresponding interpretable song feature. 

Thus, we would be able to back our recommendations in terms of simple interpretable song features.
\\
Out of the above algorithms, we went through with Collaborative Filtering. Our tests with matrix factorization were not promising due to the size of our dataset. We were unable to improve on the large computation time needed. 

\subsection{Data Workflow}
The process involved taking the spotify user id as input  and getting data output for developing tableau dashboard. It can be divided into the following steps: 1) Taking user Spotify user-id and pulling all their playlists and extracting all songs 2) Mapping user genre affinity. 3) Identify genre representative songs. 4) Creating user song matrix for a new user that is used as an input for collaborative filtering algorithm. 5) Generating personalised playlist using recommendation engine. 6) Generating recommendations from  "Top 500 songs of 2022" and "Top 500 songs of All Time" playlists by mapping ranked output to the mentioned playlist using cosine similarity. As user has intrinsic preferences towards a type of song, we exploited this assumption to map recommended songs to out of corpus songs using content based similarity. 
The data pipeline produces 3 output files that are used as an input for Tableau dashboard.
\begin{figure}[h]
  \includegraphics[width=8cm]{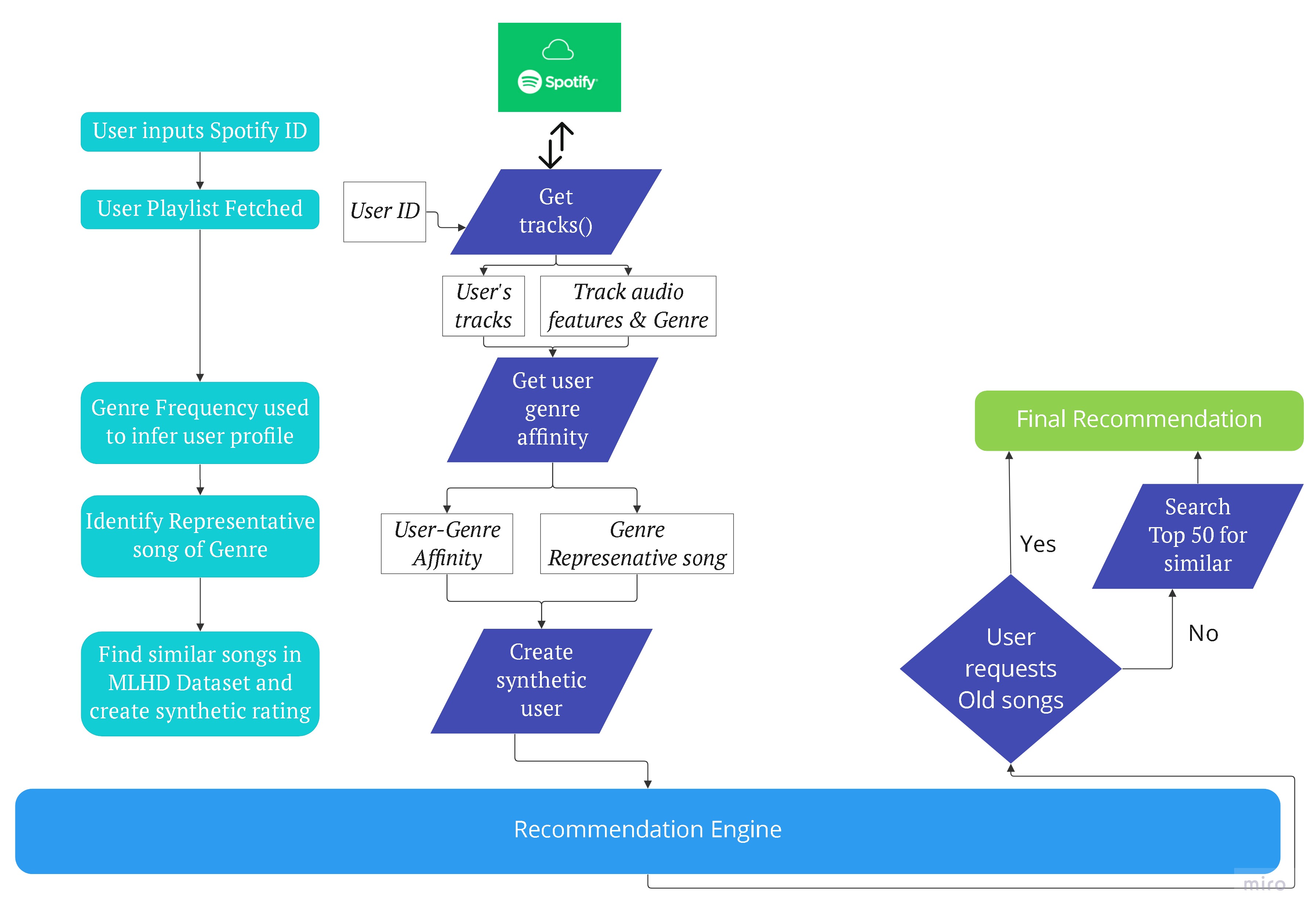}
  \caption{Data Integration Flow Chart}
\end{figure}
\subsection{Visualization and Product Features}
The final product is a Tableau dashboard that uses Data workflow output. It shows the personalised recommended playlist and graphical representation of music attributes. Further, it provide tools to control the recommendations and dynamically update the playlist as per user's mood.
\begin{figure}[h]
  \centering
  \includegraphics[width=\linewidth]{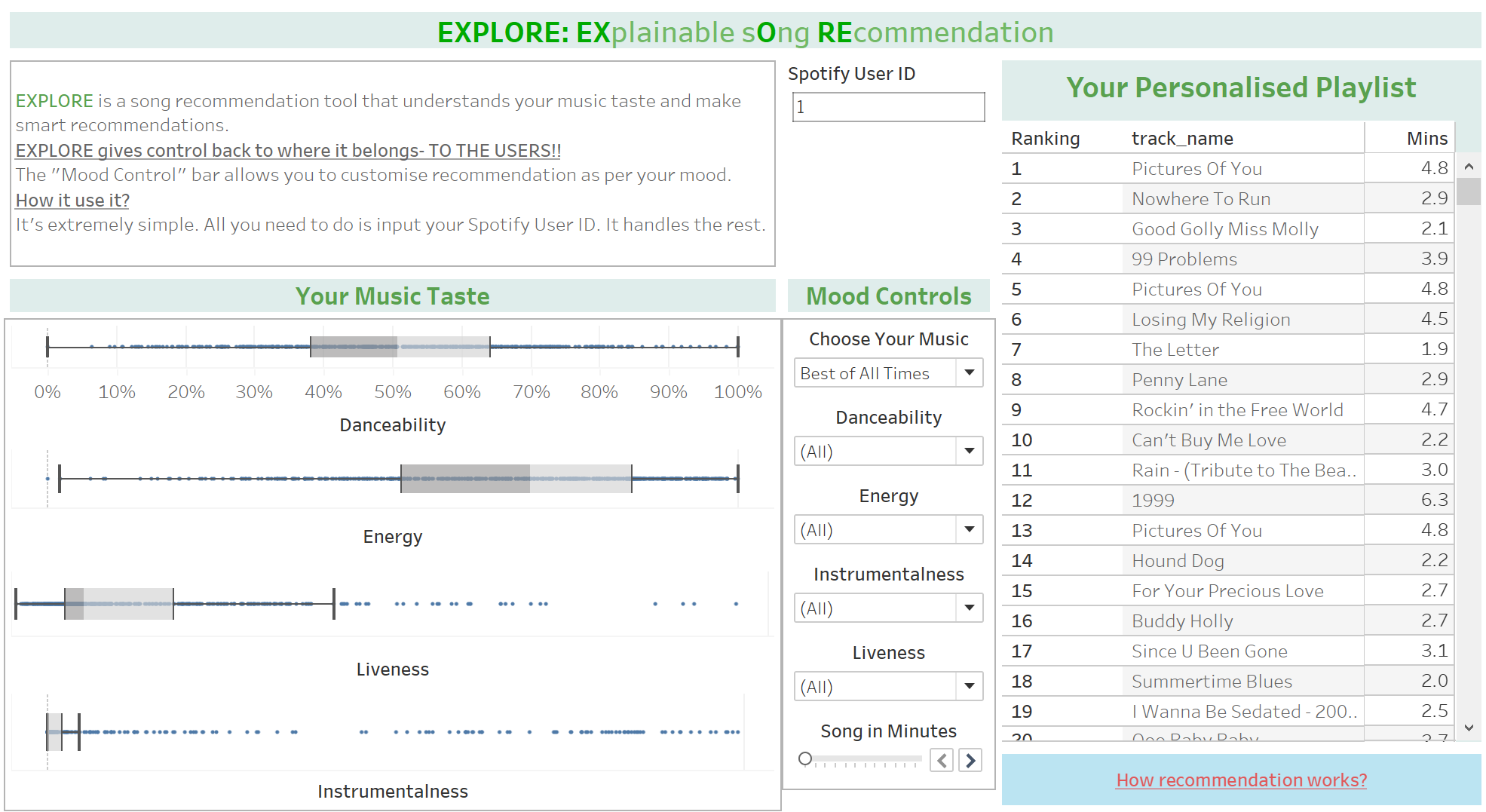}
  \caption{Product User Interface}
\end{figure}
\\
The user interface has the following sections: 1) Input field to take user's spotify id. 2) Playlist panel that has recommended songs along with their respective ranks 3) Graphical representation of recommended playlist's song attributes. 4) Mood Control bar to control and change the recommendations as per user's mood. 5) "How recommendation works?" button to describe and visualise how recommendations work.

\begin{figure}[h]
  \centering
  \includegraphics[width=2cm]{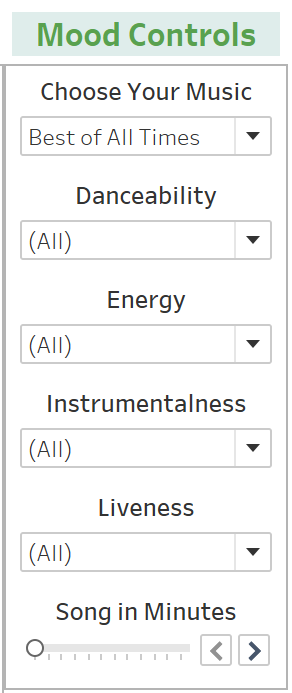}
  \caption{Mood Control Bar}
\end{figure}

The mood control section has the following toggles: 1) Choose Your Music: Selecting recommendation type from 3 options- Feeling Nostalgic, Best of 2022, and Best of All Times. 2)Danceability 3) Energy 4) Instrumentalness 5) Liveness 6) Song in Minutes: Length of a song in minutes.
\begin{figure}[h]
  \centering
  \includegraphics[width=\linewidth]{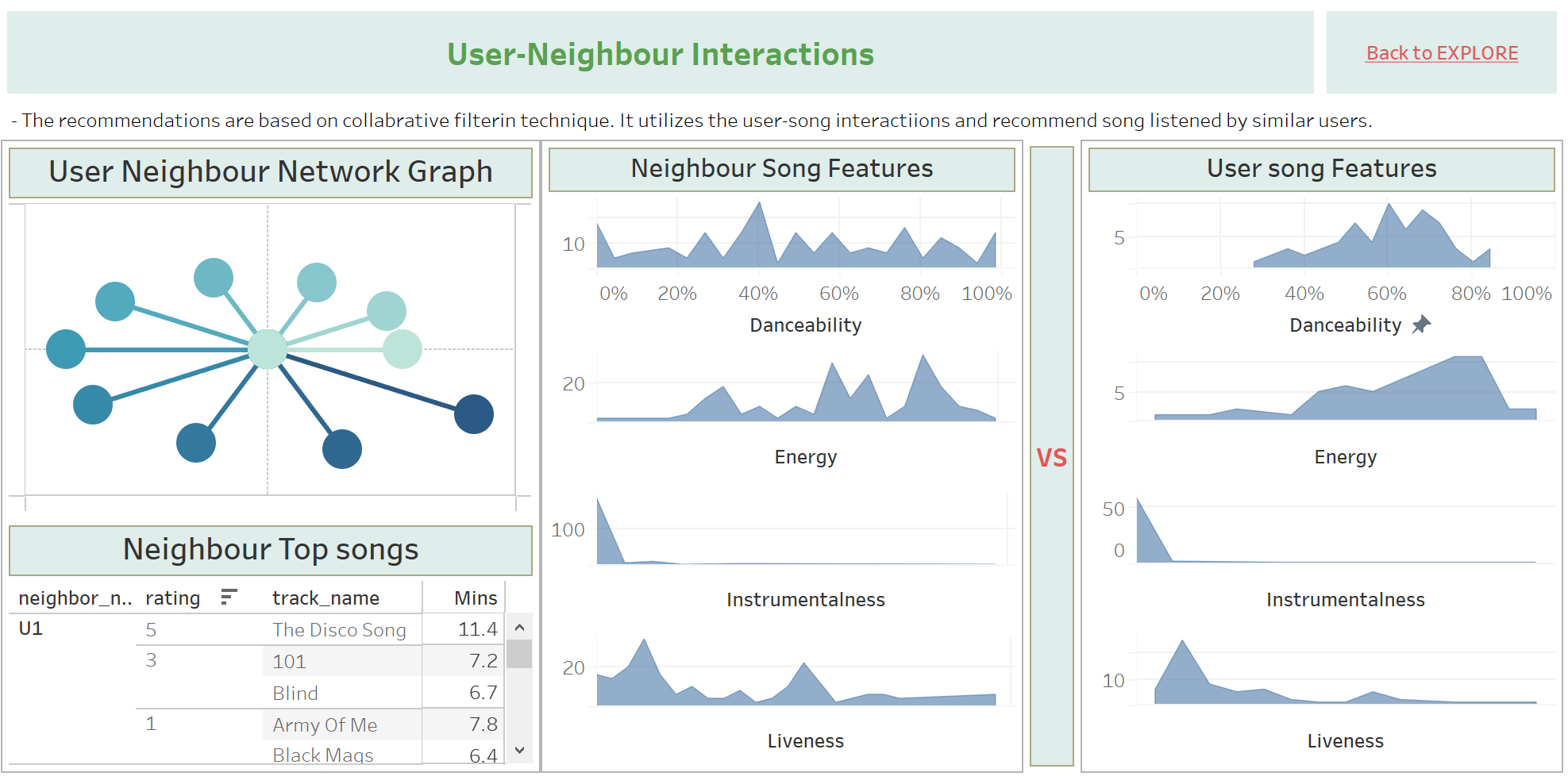}
  \caption{User Neighbour Interaction View}
\end{figure}
\\
The "How recommendation works?" button showcases the explainability aspect of the product. It visually explains the concept of collaborative filtering and visualizes user-neighbour interactions.

\subsection{Innovations}
Summarizing the above sections, we introduced following innovation through our product : 1) Defining rating matrix using song listening history of users. 2) Explainability aspect to recommendation i.e. explaining the attributes of the recommended playlist. 3) Interactive Mood controls giving users control over their recommendations. 4) Toggle to recommend songs from "Top 500 songs of 2022" and "Top 500 songs of All Time" playlists i.e. out of corpus recommendations. 5) Functionality to recommend songs to new users i.e. Spotify users not part of MLHD dataset 5) Visualizing working of recommendation algorithm using user-neighbour interactions.

\section{Evaluation}
We conduct following evaluation process for our recommendation system.
\begin{enumerate}
\item \textbf{Train-Test Validation:} In this step, we evaluate the recommendation system algorithm using train-test validation. We use the train data to develop our model using the above mentioned methodology and use it to predict the user rating for songs in test data. We use RMSE (root mean square error) metric to evaluate the performance.
$$\mathbf{RMSE} = \sqrt{\frac{1}{n}\sum\nolimits_{n=1}^N}(\epsilon_i - \hat\epsilon)^2$$
\item \textbf{User Validation:} We conducted a user study where we asked a few users to share their spotify user-id and our product generated a playlist recommendation. We then asked the user to rate the recommendation as good or bad. We defined the performance on the basis of the accuracy of recommendations.

$$\textbf{Accuracy} = \frac{\textbf{Total Good Recommendations}*100}{\textbf{Total Recommendation}}$$
\end{enumerate}

 The results were as follows: 
 
$\textbf{Ranking Based Evaluations}$\\
Although we can evaluate our recommendation systems through the above 2 methods, it is important thing when making out recommendations is the order in which recommendations are made . Thus, we will use the 2 below mentioned metrics for evaluating the output of our recommendation system . \\\\
\textbf{Ranking Evaluation Metrics:}\\
\textbf{1. Mean Average Precision@K:}\\
This metric measures how many of the recommended results are relevant and are showing at the top.
Precision@K is the fraction of relevant items in the top K recommended results. The mean average precision@K measures the average precision@K averaged over all queries in the dataset.\\
$$AP@k = \frac{1}{r_k}\sum_{k=1}^{k}s_k * rel(k)/k$$,\\
where $s_k$ is the number of relevant songs in top k results, rel(k) = 1 if kth songs are relevant and 0 otherwise
$$\boldsymbol{mAP@k = \frac{1}{N}\sum_{i=1}^{N}AP@k}$$\\
\textbf{2. Normalized Discounted Cumulative Gain:}\\
Gain refers to the relevance score for each item (song in this case) recommended. Cumulative gain at K is the sum of gains of the first K items recommended. Discounted cumulative gain (DCG) weighs each relevance score based on its position, with a higher weight to the recommendations at the top. \\
$$NDCG@K = \frac{DCG@K}{IDCG@K}$$, where\\
$\displaystyle{DCG@K = \sum_{i=1}^{K} \frac{G_i}{log_2{(i+1)}}}$ and\\ $\displaystyle{IDCG@K = \sum_{i=1}^{K_i} \frac{G_i}{log_2{(i+1)}}}$, where $K_i$ is K(ideal) and $G_i$ is G (relevance score of each recommended item). 
\\\\
\textbf{Results:}\\
We followed two different approaches to sample our dataset into 80\%-20\% train-test split. The first approach takes a stratified sampling route where for each user the split is done randomly into 80-20 split. The second approach does an overall random shuffling and we obtain the train (80\%) and test (20\%).

\begin{figure}[h]
  \centering
  \includegraphics[width=\linewidth]{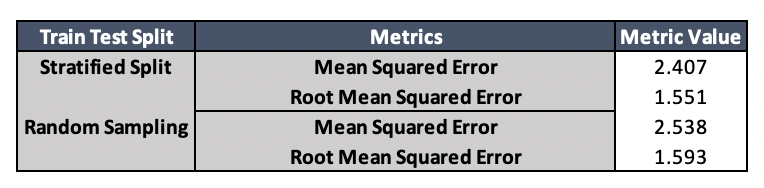}
  \caption{RMSE-MSE Evaluation}
\end{figure}

\begin{figure}[h]
  \centering
  \includegraphics[width=\linewidth]{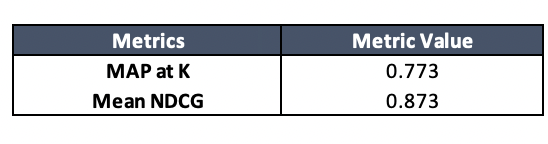}
  \caption{Ranking Evaluation}
\end{figure}

The ranking metrics (MAP\@K for k = 3 : 0.773, Mean NDCG: 0.873) and the RMSE values show that our collaborative filtering algorithm does a good job in the prediction of ratings as well as generating relevant song recommendations.\\

\section{Experiments}
We conducted user survey to seek feedback on our product. We got 17 responses and the overall response was positive. One promising observation has been a strong positive response for "Explainability Visualization" from respondents who did not care about explainability in the first place.

\begin{figure}[h]
  \includegraphics[width=\linewidth]{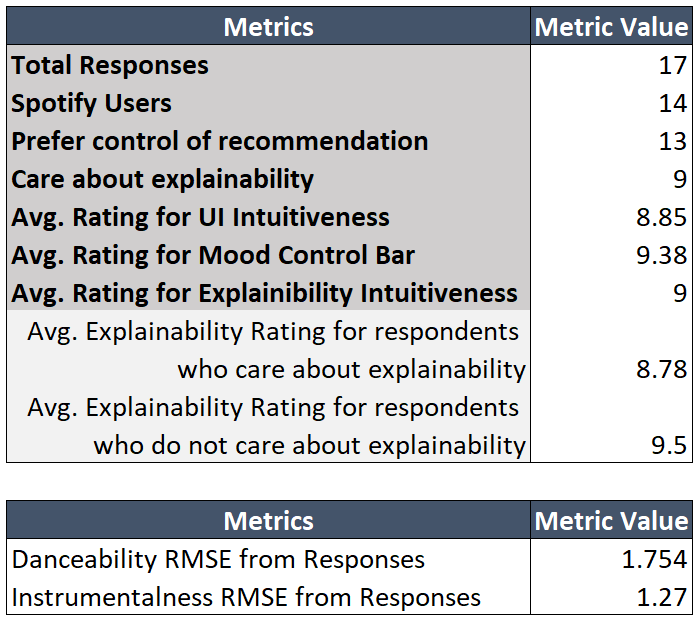}
  \caption{User Survey Evaluation}
\end{figure}

\section{Conclusion and Future Scope}
Through our final product, the team was able to address all the key challenges mentioned in the "Problem Definition" section. Following milestones were achieved: 1) Product to generate personalised playlists. 2) Incorporate explainability aspect to recommendations. 3) Adding tools for users to control and dynamically update recommendations.\\
However, there are shortcomings which we would like to work in future. 1) The product might face challenges with scalability due to a) dependency on Spotify API to fetch songs and b) recommendation algorithm latency 2) Recommendation is based on old data and on the assumption that music tastes are purely based on song characteristics.  

Although we are answering a few existing challenges in state-of-the-art recommendation systems, there is potential to improve on the following aspects in future: 1) Making recommendations for a group. 2) Incorporating user feedback for recommended songs in UI( Like/Dislike button) and using their responses to improve the recommendation.

\bibliographystyle{ACM-Reference-Format}
\bibliography{references}










\end{document}